\documentclass[aps,prb,twocolumn]{revtex4-1}   % style for Physical Review B and AJP are similar

\usepackage{amsmath}    % need for subequations
\usepackage{amsfonts}  %note how statements can be commented out
\usepackage{amssymb}
\usepackage{graphicx}   % for figures

\usepackage[latin1]{inputenc}     % pour les accents
\usepackage[T1]{fontenc}

\usepackage{float}   % pour que les images soient inclusent dans le texte
\usepackage{color}   % pour faire des surlignes de couleur
\usepackage{soul}    % pour faire des surlignes de couleur
\graphicspath{{./images/}}

\newtheorem{Theoreme}{THEOREM}
\newtheorem{Definition}{Definition}

\definecolor{lightblue}{rgb}{0.8,0.9,1} % bleu ciel

\begin{document}

\title{The theory of the double preparation: discerned and indiscerned particles}
\author{Michel Gondran}
 %\altaffiliation[Also at ]{home.}  %  optional
 \affiliation{University
Paris Dauphine, Lamsade, 75 016 Paris, France}
 \email{michel.gondran@polytechnique.org}   %optional
\author{Alexandre Gondran}
\affiliation{\'Ecole Nationale de l'Aviation Civile, 31000 Toulouse,
France}
 \email{alexandre.gondran@enac.fr}   %optional

\begin{abstract}

In this paper we propose a deterministic and realistic quantum
mechanics interpretation which may correspond to Louis de Broglie's "double
solution theory". Louis de
Broglie considers two solutions to the Schr\"odinger equation, a
singular and physical wave u representing the particle (soliton
wave) and a regular wave   representing probability (statistical
wave). We return to the idea of two solutions, but in the form of
an interpretation of the wave function based on two different
preparations of the quantum system.
We demonstrate the necessity of 
this double interpretation when the particles are subjected to a
semi-classical field by studying the convergence of the
Schr\"odinger equation when the Planck constant tends to 0. 
For this convergence, we reexamine not only the foundations of quantum mechanics but also 
those of classical mechanics, and in particular two important paradox of classical mechanics: 
the interpretation of the principle of least action and the 
the Gibbs paradox. 
%We show that it is natural 
%to introduce the concepts of discerned and indiscerned particles both in classical %mechanics and in quantum mechanics. 
We find
two very different convergences which depend on the preparation of
the quantum particles: particles called indiscerned (prepared in
the same way and whose initial density is regular, such as atomic
beams) and particles called discerned (whose density is singular,
such as coherent states). These results are based on the Minplus
analysis, a new branch of mathematics that we have developed
following Maslov, and on the Minplus path integral which is the
analog in classical mechanics of the Feynman path integral in
quantum mechanics. The indiscerned (or discerned) quantum
particles converge to indiscerned (or discerned) classical
particles and we deduce that the de Broglie-Bohm pilot wave is the
correct interpretation for the indiscerned quantum particles
(wave statistics) and the Schr\"odinger interpretation is the
correct interpretation for discerned quantum particles (wave
soliton). Finally, we show that this double interpretation can be extended to the non semi-classical case.

\end{abstract}

\maketitle

%%%%%%%%%%%%%%%%%%%%%%%%%%%%%%%%%%%%%%%%%%%%%%%%%%%%%%%%%%%%%%%%%%%%%%%%%%%%%%
%%%%%%%%%%%%%%%%%%%%%%%%%%%%%%%%%%%%%%%%%%%%%%%%%%%%%%%%%%%%%%%%%%%%%%%%%%%%%%
%%%%%%%%%%%%                    INTRODUCTION                     %%%%%%%%%%%%%
%%%%%%%%%%%%%%%%%%%%%%%%%%%%%%%%%%%%%%%%%%%%%%%%%%%%%%%%%%%%%%%%%%%%%%%%%%%%%%
%%%%%%%%%%%%%%%%%%%%%%%%%%%%%%%%%%%%%%%%%%%%%%%%%%%%%%%%%%%%%%%%%%%%%%%%%%%%%%
\section{Introduction}
\label{sect:intro}

For Louis de Broglie, the correct interpretation of quantum mechanics
was the "theory of the double solution" introduced in
1927~\cite{Broglie1927} and for which the pilot-wave was just a
low-level product~\cite{Broglie}: \textit{I introduced as a 'double solution theory' the idea that
it was necessary to distinguish two different solutions but both
linked to the wave equation, one that I called wave $u$ which was
a real physical wave but not normalizable having a local anomaly
defining the particle and represented by a singularity, the other
one as the Schr\"odinger $\Psi$ wave, which is normalizable
without singularities and being a probability representation.}

Louis de Broglie distinguishes two solutions to the Schr\"odinger
equation, a singular and physical wave $u$ representing the
particle (soliton wave) and a regular wave $\Psi$ representing
probability (statistical wave). But, de Broglie don't have
never find a consistent "double solution theory". We return to the idea of two solutions, but in the form of a
double interpretation of the wave function based on different
preparations of the quantum system.
We demonstrate the necessity of this double
interpretation when the particles are subjected to a
semi-classical field by studying the convergence of the
Schr\"odinger equation when the Planck constant tends to
0~\cite{Gondran2011,Gondran2012a}. This convergence of quantum to classical mechanics 
poses three types of difficulty that seem insurmountable:
\begin{itemize}
 \item a physical difficulty, because h is a constant and therefore its convergence to 0 is not physical;
 \item a conceptual difficulty: in quantum mechanics, particles are regarded as indistinguishable 
whereas they are considered to be distinguishable in classical mechanics;
 \item mathematical difficulties of convergence of equations.
\end{itemize}
The physical difficulty is the easiest to solve: it is only mathematically, not physically, that we decrease 
the Planck constant to 0; numerically we obtain the same results if we increase the mass $m$ of the particle to infinity. 

The conceptual difficulty forces us into reexamining not only the foundations of quantum mechanics but also 
those of classical mechanics. It is necessary to understand and solve two important paradoxes of classical mechanics: 
the interpretation of the principle of least action where the "final causes" seem to be substituted for the "efficient causes";  
the Gibbs paradox where the entropy calculation of a mixture of two identical gases by classical mechanics with distinguishable
particles leads to an entropy twice as big as expected. We solve the conceptual difficulty by showing that it is natural 
to introduce the concepts of discerned and indiscerned particles both in classical mechanics and in quantum mechanics. 

The mathematical difficulties will be greatly simplified by considering two types of initial conditions (discerned and 
indiscerned particles) which yield very different mathematical convergences. They are also simplified by the Minplus
analysis~\cite{Gondran_1996, GondranMinoux}, a new branch of mathematics that we have developed
following Maslov~\cite{Maslov,Maslov2}. 
The paper is organized as follows. In section~\ref{sect:ELHJ}, we will show that the difficulties of interpretation of the
principle of least action concerning the "final causes" come from the "Euler-Lagrange action" (or classical action)
$S_{cl}(\textbf{x},t;\textbf{x}_0)$, which links the initial
position $\textbf{x}_0$ and its position $\textbf{x}$ at time t
and not from the "Hamilton-Jacobi action" $S(\textbf{x},t)$, which depends
on an initial action $S_{0}(\textbf{x})$. These two actions are solutions to the same Hamilton-Jacobi
equation, but with very different initial conditions: smooth
conditions for the Hamilton-Jacobi action, singular conditions for
the Euler-Lagrange action. 

In section~\ref{sect:Minplus}, we show how Minplus analysis, a new branch of
nonlinear mathematics, explains the difference between the
Hamilton-Jacobi action and the Euler-Lagrange action. We obtain
the equation between these two actions, which we call the Minplus
path integral: it is the
analog in classical mechanics of the Feynman path integral in
quantum mechanics. We show that it is the key to
understanding the principle of least action.

In section~\ref{sect:Discerned}, we introduce in classical
mechanics the concept of indiscerned particles through the statistical Hamilton-Jacobi equations. The discerned particles in classical
mechanics correspond to a
deterministic action $S(\textbf{x},t;\textbf{x}_0,\textbf{v}_0)$,
which links a particle in initial position $\textbf{x}_0$ and
initial velocity $\textbf{v}_0$ to its position $\textbf{x}$ at
time $t$ and verifies the deterministic Hamilton-Jacobi equations.
And the Gibbs paradox is solved by the indiscerned  particles in classical mechanics.

In section~\ref{sect:2limits}, we study the convergence of quantum mechanics to classical
mechanics when the Planck constant tends to 0 by considering two cases : the first corresponds
to the convergence to an indiscerned classical particle, and the
second corresponds to the convergence to a classical discerned particle.~\cite{Gondran2011,Gondran2012a}
Based on these convergences, we propose a new interpretation of
quantum mechanics, the "theory of the double preparation", a response that corresponds to Louis de Broglie's "theory of the double solution".

In section~\ref{sect:NonSemiClassical}, we generalize this interpretation when the 
semi-classical approximation is not valid. Following de Muynck~\cite{Muynck}, we show that it
is possible to construct a deterministic field quantum theory that
extends the previous double semi-classical interpretation to the
non semi-classical case.

\section{The Euler-Lagrange and Hamilton-Jacobi actions}
\label{sect:ELHJ}

The intense debate on the interpretation of the wave function in quantum 
mechanics for eighty years has in fact left the debate on the interpretation 
of the action and the principle of least action in classical mechanics in the dark, 
since their introduction in 1744 by Pierre-Louis Moreau de Maupertuis \cite{Maupertuis1744}: 
"\emph{Nature, in the
production of its effects, does so always by the simplest means
[...] the path it takes is the one by which the quantity of action
is the least}". Maupertuis understood that, under certain conditions, Newton's 
equations are equivalent to the fact that a quantity, which he calls the action, 
is minimal. Indeed, one can verify that the trajectory realized in Nature is that 
which minimizes (or renders extremal) the action, which is a function depending 
on the different possible trajectories.

However, this
principle has often been viewed as puzzling by many scholars,
including Henri Poincar\'e, who was nonetheless one of its most
intensive users \cite{Poincare}: "\emph{The very enunciation of
the principle of least action is objectionable. To move from one
point to another, a material molecule, acted on by no force, but
compelled to move on a surface, will take as its path the geodesic
line, i.e., the shortest path. This molecule seems to know the
point to which we want to take it, to foresee the time it will
take to reach it by such a path, and then to know how to choose
the most convenient path. The enunciation of the principle
presents it to us, so to speak, as a living and free entity. It is
clear that it would be better to replace it by a less
objectionable  enunciation, one in which, as philosophers would
say, final effects do not seem to be substituted for acting
causes.}"

We will show that the difficulties of interpretation of the
principle of least action concerning the "final causes" or the
"efficient causes" come from the existence of two different
actions: the "Euler-Lagrange action"
$S_{cl}(\textbf{x},t;\textbf{x}_0)$
and the "Hamilton-Jacobi action" $S(\textbf{x},t)$.

Let us consider a system evolving from the position
$\textbf{x}_{0}$ at initial time to the position $\textbf{x}$ at
time $t$ where the variable of control \textbf{u}(s) is the
velocity:
\begin{eqnarray}\label{eq:evolution}
\frac{d \textbf{x}\left( s\right) }{ds}=\mathbf{u}(s),\qquad\forall s\in\left[ 0,t\right]\\
\label{eq:condinitiales}
\textbf{x}(0) =\mathbf{x}_{0},\qquad\textbf{x}(t) =\mathbf{x}.
\end{eqnarray}

If $L(\textbf{x},\dot{\textbf{x}},t)$ is the Lagrangian of the
system, when the two positions $\textbf{x}_0$ and $\textbf{x}$ are
given, \emph{the Euler-Lagrange action} $S_{cl}(\mathbf{x},t;
\textbf{x}_0) $ is the function defined by:
\begin{equation}\label{eq:defactioncondit}
S_{cl}(\textbf{x},t;\mathbf{x}_{0})=\min_{\textbf{u}\left(
s\right),0 \leq s\leq t} \int_{0}^{t}L(\textbf{x}(s),
\textbf{u}(s),s)ds,
\end{equation}
where the minimum (or more generally an extremum) is taken on the
controls $\mathbf{u}(s)$, $s\in$ $\left[ 0,t\right]$, with the
state $\textbf{x}(s)$ given by equations (\ref{eq:evolution}) and
(\ref{eq:condinitiales}). This is the principle of least action
defined by Euler \cite{Euler1744} in 1744 and Lagrange \cite{Lagrange} in 1755. 

The solution $(\widetilde{\textbf{x}}(s),\widetilde{\textbf{u}}(s)
)$ of (\ref{eq:defactioncondit}), if the Lagrangian
$L(\textbf{x},\dot{\textbf{x}},t) $ is twice differentiable,
satisfies the Euler-Lagrange equations on the interval $[0,t] $:
\begin{equation}\label{eq:EulerLagrange1}
\frac{d}{ds}\frac{\partial L}{\partial
\dot{\textbf{x}}}(\textbf{x}(s),\dot{\textbf{x}}(s),s)-
\frac{\partial L}{\partial
\textbf{x}}(\textbf{x}(s),\dot{\textbf{x}}(s),s)=0,\quad\forall s\in\left[ 0,t\right]%~~~~~~~~~~(0\leqs \leq t)
\end{equation}
\begin{equation}\label{eq:EulerLagrange12}
\textbf{x}(0) =\mathbf{x}_{0},~~~~\textbf{x}(t) =\mathbf{x}.
\end{equation}
For a non-relativistic particle in a linear potential field with
the Lagrangian $L(\mathbf{x},\mathbf{\dot{x}},t)= \frac{1}{2}m
\mathbf{\dot{x}}^2 + \textbf{K}. \textbf{x}$, equation
(\ref{eq:EulerLagrange1}) yields $\frac{d}{ds}( m
\dot{\textbf{x}}(s)) - \textbf{K}=0 $. The trajectory minimizing
the action is $\widetilde{\textbf{x}}(s)= \textbf{x}_0 +
\frac{s}{t} (\textbf{x} -\textbf{x}_0)- \frac{\textbf{K}}{2 m} t s
+ \frac{\textbf{K}}{2 m} s^2$, and the Euler-Lagrange action is
equal to
\begin{equation}\label{eq:actionlinearEulerLagrange}
S_{cl}( \mathbf{x},t; \textbf{x}_0)= m
\frac{(\textbf{x}-\textbf{x}_0)^2}{2 t}+ \frac{K .(\textbf{x} +
\textbf{x}_0)}{2}t - \frac{K^2}{24 m}t^3.
\end{equation}

\begin{figure}[h]
\includegraphics[width=0.8\linewidth]{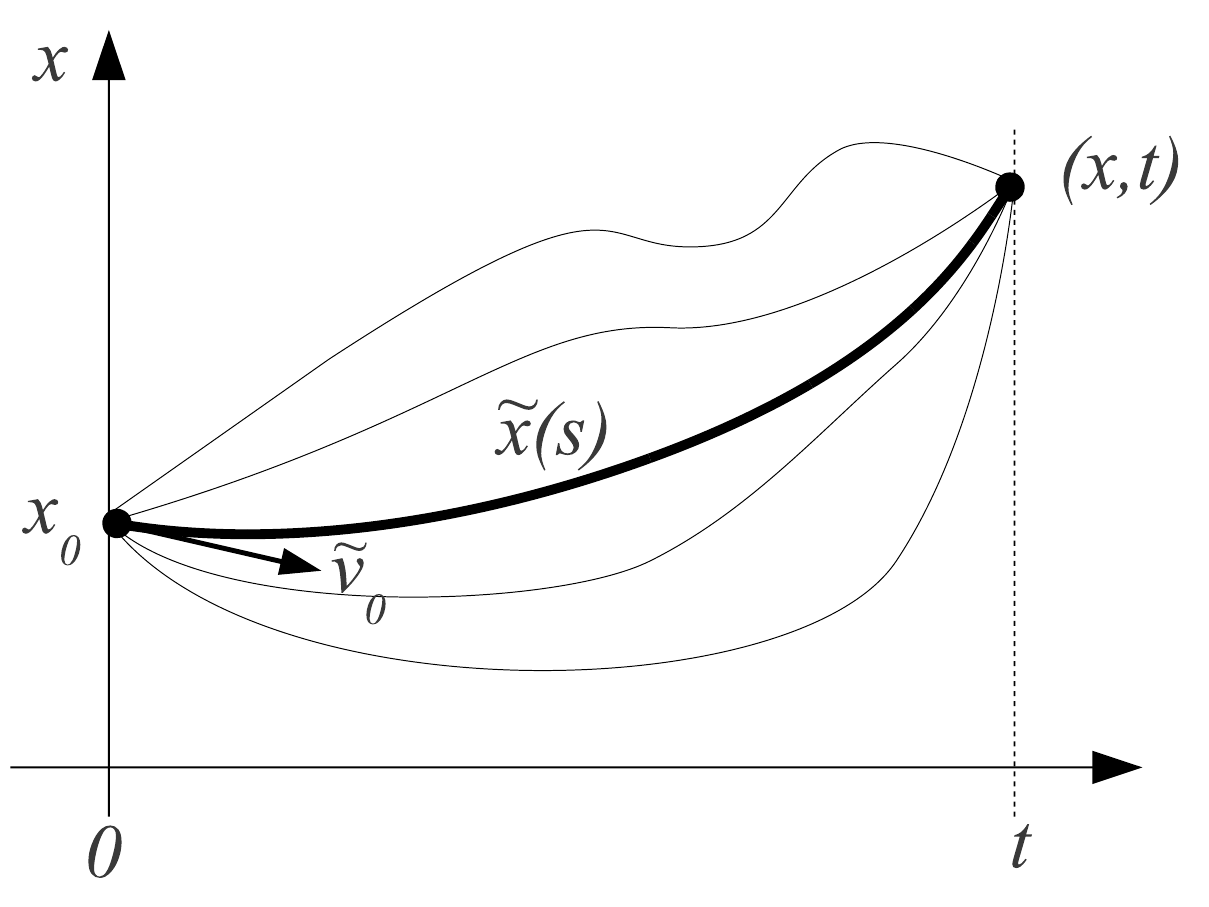}
\caption{\label{fig:trajEL} Different trajectories $x(s) $ ($0\leq
s \leq t$) between $(x_0,0)$ and $(x,t)$ and the optimal trajectory $\widetilde{x}(s)$ 
with $\widetilde{\textbf{v}}_0= \frac{\textbf{x}-\textbf{x}_0}{t}-\frac{K t}{2m}$.}
\end{figure}

Figure \ref{fig:trajEL} shows different trajectories going from $x_0$ at time 
$t=0$ to $x$ at final time $t$. The parabolic trajectory $\widetilde{x}(s) $ 
corresponds  to this which realizes the minimum in the equation (\ref{eq:defactioncondit}).

Equation (\ref{eq:defactioncondit}) seems to show that, among the
trajectories which can reach ($\textbf{x},t$) from the initial
position $\textbf{x}_0 $, the principle of least action allows to
choose the velocity at each time. In reality, the principle of
least action used in equation (\ref{eq:defactioncondit}) does not
choose the velocity at each time $s$ between $0$ and $t$, but only when
it arrives at $\textbf{x}$ at time $t$. The knowledge
of the velocity at each time $s$ ($0\leq
s \leq t$) requires the resolution of the
Euler-Lagrange equations
(\ref{eq:EulerLagrange1},\ref{eq:EulerLagrange12}) on the whole
trajectory. In the case of a non-relativistic particle
in a linear potential field, the velocity at time $s$ ($0\leq
s \leq t$) is $\widetilde{\textbf{v}}(s)= \frac{\textbf{x}-\textbf{x}_0}{ t} - \frac{K t}{2 m}+\frac{K s}{ m}$ with the initial velocity
\begin{equation}\label{eq:vitesseEulerinit}
\widetilde{\textbf{v}}_0= \frac{\textbf{x}-\textbf{x}_0}{ t}- \frac{K t}{2 m}.
\end{equation}
Then, $\widetilde{\textbf{v}}_0$ depends on the position $\textbf{x}$ of the particle at the final
time $t$. This dependence of the "final causes" is
general. This is the Poincar\'e's main criticism of the principle of
least action: "\emph{This molecule seems to know the point to
which we want to take it, to foresee the time it will take to
reach it by such a path, and then to know how to choose the most
convenient path.}"

One must conclude that, without knowing the initial velocity, the
Euler-Lagrange action answers a problem posed by an observer, and
not by Nature: "What would be the velocity of the particle at the
initial time to arrive in $\textbf{x}$ at time $t$?" The resolution
of this problem implies that the observer solves the
Euler-Lagrange equations
(\ref{eq:EulerLagrange1},\ref{eq:EulerLagrange12}) after the
observation of $\textbf{x}$ at time $t$. This is an \emph{a
posteriori} point of view.

But from 1830, Hamilton~\cite{Hamilton} proposes to consider the action $S$ as a
function of the coordinates and of the time $S(\textbf{x},t)$. It
is customary to call it Hamilton's principal function\cite{Landau,Goldstein,Butterfield}. In the following, we refer
to it as the Hamilton-Jacobi action.  Indeed, for the Lagrangian $L(\mathbf{x},\mathbf{\dot{x}},t)= \frac{1}{2}m
\mathbf{\dot{x}}^2 - V(\textbf{x},t)$, this action
satisfies the Hamilton-Jacobi equations:
\begin{eqnarray}\label{eq:HJ}
\frac{\partial S}{\partial t}+\frac{1}{2m}(\triangledown
S)^{2}+V(\textbf{x},t)=0
\\\label{eq:condinitialHJ}
S(\textbf{x},0)=S_{0}(\textbf{x}).
\end{eqnarray}

The initial condition $S_{0}(\textbf{x}) $ is essential to
defining the general solution to the Hamilton-Jacobi equations
(\ref{eq:HJ},\ref{eq:condinitialHJ}) although it is ignored in
the classical mechanics textbooks such as those of Landau \cite{Landau} chap.7 §
47 and Goldstein \cite{Goldstein}  chap. 10.
However, the initial condition $S_{0}(\textbf{x}) $ is
mathematically necessary to obtain the general solution to the
Hamilton-Jacobi equations. Physically, it is the condition that
describes the preparation of the particles. We will see that this
initial condition is the key to understanding the principle of
least action.

The main property of the Hamilton-Jacobi action is that the velocity of a
non-relativistic classical particle is given for each point
$\left( \mathbf{x,}t\right)$ by:
\begin{equation}\label{eq:eqvitesse}
\mathbf{v}\left( \mathbf{x,}t\right) =\frac{\mathbf{\nabla }S\left( \mathbf{%
x,}t\right) }{m}.
\end{equation}

In the general case where $S_0(\textbf{x})$ is a regular function,
for example differentiable, equation (\ref{eq:eqvitesse}) shows
that the solution $S\left( \mathbf{x,}t\right) $ of the
Hamilton-Jacobi equations yields the velocity field  for each
point ($\textbf{x},t$) from the velocity field $\frac{\nabla
S_0(\textbf{x})}{m} $ at the initial time. In particular, if at
the initial time, we know the initial position $\textbf{x}_{init}$
of a particle, its velocity at this time is equal to $\frac{\nabla
S_0(\textbf{x}_{init})}{m}$. From the solution $ S\left(
\mathbf{x,}t\right)$ of the Hamilton-Jacobi equations, we deduce
with (\ref{eq:eqvitesse}) the trajectories of the particle. The
Hamilton-Jacobi action $S\left( \mathbf{x,}t\right)$ is then a
field which "pilots" the particle.

%%%%%%%%%%%%%%%%%%%%%%%%%%%%%%%%%%%

There is another solution to the Hamilton-Jacobi equation; it is
the Euler-Lagrange action. Indeed,
$S_{cl}(\textbf{x},t;\textbf{x}_0)$ satisfies the Hamilton-Jacobi
(\ref{eq:HJ}) with the initial condition
\begin{equation}\label{eq:conditinitEL}
S(\textbf{x},0)= \{ 0~\text{ if }~\textbf{x}=\textbf{x}_0,+\infty~\text{ if not} \}
\end{equation}
which is a very singular function. Mathematical analysis will help us to interpret the solution
to the Hamilton-Jacobi equations and the principle of least
action.

\section{Minplus analysis and the Minplus path integral}
\label{sect:Minplus}

There exists a new branch of mathematics, Minplus analysis,
which studies nonlinear problems through a linear approach, cf.
Maslov~\cite{Maslov,Maslov2} and
Gondran~\cite{Gondran_1996,GondranMinoux}. The idea is to
substitute the usual scalar product $\int_{X} f(x) g(x) dx$ with
the Minplus scalar product:
\begin{equation}
    (f,g) =\inf_{x\in X}\left\{ f(x)+g(x) \right\}
\end{equation}
In the scalar product we replace the field of the real number $(
\mathbb{R},+,\times )$ with the algebraic structure \textit{Minplus} $(
\mathbb{R}\cup \{ +\,\infty \} ,\min ,+)$, i.e. the set of real numbers
(with the element infinity $\{ +\infty \}$) endowed with the
operation Min (minimum of two reals), which remplaces the usual
addition, and with the operation + (sum of two reals), which
remplaces the usual multiplication. The element $\{+\,\infty \}$
corresponds to the neutral element for the operation Min, Min$( \{
+\infty \} ,a) =a$ $\forall a\in \mathbb{R}$. This approach bears a close
similarity to \emph{the theory of distributions for the nonlinear
case}; here, the operator is "linear" and continuous with respect
to the Minplus structure,
though \emph{nonlinear} with respect to the classical structure $%
\left( \mathbb{R},+,\times \right)$. In this Minplus structure, the
Hamilton-Jacobi equation is linear, because if $S_1(\textbf{x},t)$
and $S_2(\textbf{x},t)$ are solutions to (\ref{eq:HJ}), then
$\min\{\lambda + S_1(\textbf{x},t), \mu + S_2(\textbf{x},t)\}$ is
also a solution to the Hamilton-Jacobi equation (\ref{eq:HJ}).

The analog to the Dirac distribution $\delta(\textbf{x})$ in
Minplus analysis is the nonlinear distribution
$\delta_{\min}(\textbf{x})=\{ 0~\text{ if }~\textbf{x}=\textbf{0}, 
+\infty~\text{ if not}\}$. With this nonlinear Dirac distribution, we can define
elementary solutions as in classical distribution theory. In
particular, we obtain:

\emph{The classical Euler-Lagrange action
$S_{cl}(\textbf{x},t;\textbf{x}_0)$ is the elementary solution to
the Hamilton-Jacobi equations
(\ref{eq:HJ})(\ref{eq:condinitialHJ}) in the Minplus analysis with
the initial condition}
\begin{equation}
S(\textbf{x},0)= \delta_{\min}(\textbf{x}- \textbf{x}_0)= \{
0~\text{ if }~\textbf{x}=\textbf{x}_0,~+\infty~\text{ if not} \}.
\end{equation}
The Hamilton-Jacobi action $S(\textbf{x},t)$ is then given by the
Minplus integral
\begin{equation}\label{eq:valactionglobale}
S(\textbf{x},t)=\inf_{\textbf{x}_0} \{ S_0(\textbf{x}_0)
 + S_{cl}(\textbf{x},t;\textbf{x}_0)\}.
\end{equation}
that we call the Minplus path integral. It is an equation similar to the Hopf-Lax formula\cite{Lions,Evans}. This equation is 
in analogy with the solution to the heat transfer equation given
by the classical integral:
\begin{equation}
        S(x,t)=\int S_{0}( x_0) \frac{1}{2\sqrt{\pi t}} e^{-\frac{
        \left( x-x_0\right) ^{2}}{4t}}dx_0,
\end{equation}
which is the product of convolution of the initial condition
$S_{0}(x) $ with the elementary solution to the heat transfer
equation $ e^{-\frac{x^{2}}{4t}}$.

This Minplus path integral yields a very simple relation between
the Hamilton-Jacobi action, the general solution to the
Hamilton-Jacobi equation, and the Euler-Lagrange actions, the
elementary solutions to the Hamilton-Jacobi equation. We can also consider that the Minplus integral
(\ref{eq:valactionglobale}) for the action in classical mechanics
is analogous to the Feynmann path integral for the wave function
in quantum mechanics. Indeed, in the Feynman paths integral \cite{Feynman_1965} (p. 58),
the wave function $\Psi(\textbf{x},t)$ at time $t$ is written as a
function of the initial wave function $\Psi_{0}(\textbf{x})$:
\begin{equation}\label{eq:interFeynman}
\Psi(\textbf{x},t)= \int F(t,\hbar)
\exp\left(\frac{i}{\hbar}S_{cl}(\textbf{x},t;\textbf{x}_{0}\right)
\Psi_{0}(\textbf{x}_{0})d\textbf{x}_0
\end{equation}
where $F(t,\hbar)$ is an independent function of $\textbf{x}$ and
of $\textbf{x}_{0}$.

For a particle in a linear potential $V(\textbf{x})= - \textbf{K}
.\textbf{x}$ with the initial action $S_0(\textbf{x})= m
\textbf{v}_0 \cdot \textbf{x}$, we deduce from equation
(\ref{eq:valactionglobale}) that the Hamilton-Jacobi action is
equal to $ S\left( \mathbf{x},t\right)=m \textbf{v}_0 \cdot
\textbf{x} - \frac{1}{2} m \textbf{v}_0^2 t +\textbf{K}.\textbf{x}
t - \frac{1}{2} \textbf{K}.\textbf{v}_{0} t^{2} -
\frac{\textbf{K}^2 t^3}{6 m}$.

\begin{figure}[h]
\includegraphics[width=0.8\linewidth]{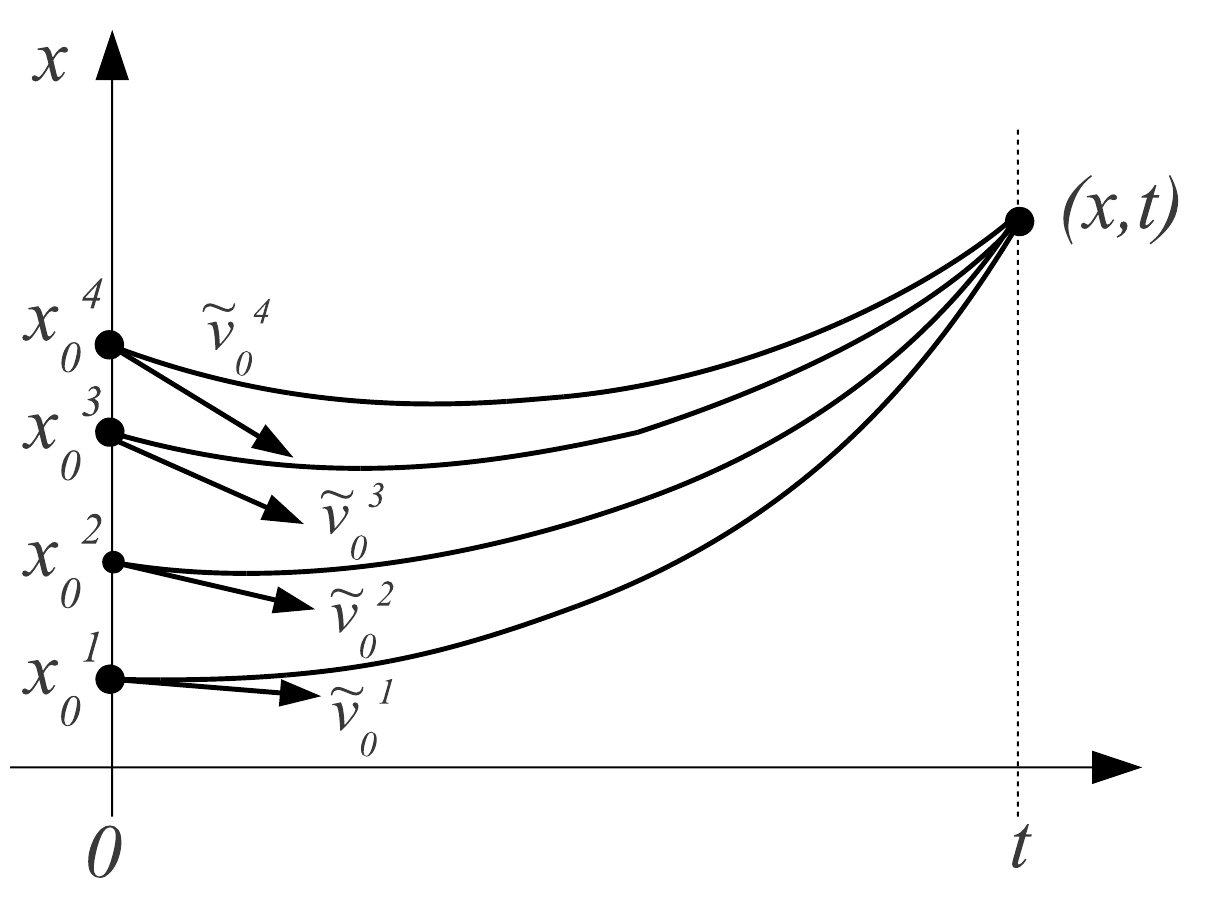}
\caption{\label{fig:trajHJ} Classical trajectories $\widetilde{x}(s) $ ($0\leq
s \leq t$) between the different initial positions $x^i_0$ and the position $x$ at time $t$. 
We obtain $\widetilde{\textbf{v}}^i_0= \frac{\textbf{x}-\textbf{x}^i_0}{ t}- \frac{K t}{2 m}$.}
\end{figure}

Figure \ref{fig:trajHJ} shows the classical trajectories (parabols) 
going from different starting points $x^i_0$ at time $t=0$ to the point $x$ at final time $t$. 
The Hamilton-Jacobi action is compute with these trajectories in the Minplus path integral (\ref{eq:valactionglobale}).

%%%%%%%%%%%%%%%%%%%%%%%%%%%%%%%%%%%

Finally, we can write the Minplus paths integral as follows:
\begin{equation}\label{eq:defactionHJ}
S(\mathbf{x},t)=\min_{\textbf{x}_0;\mathbf{u}\left( s\right),0
\leq s\leq t }\left\{ S_{0}\left( \mathbf{x}_{0}\right)
+\int_{0}^{t}L(\textbf{x}(s), \mathbf{u}(s),s)ds\right\}
\end{equation}
where the minimum is taken on all initial positions $\textbf{x}_0$
and on the controls $\mathbf{u}(s)$, $s\in$ $\left[ 0,t\right]$,
with the state $\textbf{x}(s)$ given by equations
(\ref{eq:evolution}) and (\ref{eq:condinitiales}). This is
possible because $S_0(\textbf{x}_0)$ does not play a role in
(\ref{eq:defactionHJ}) for the minimization on $\textbf{u}(s)$.
 
Equation (\ref{eq:defactionHJ}) seems to show that, among the
trajectories which can reach ($\textbf{x},t$) from an unknown
initial position and a known initial velocity field, Nature
chooses the initial position and at each time the velocity that
yields the minimum (or the extremum) of the Hamilton-Jacobi
action.

Equations (\ref{eq:eqvitesse}), (\ref{eq:HJ}) and
(\ref{eq:condinitialHJ}) confirm this interpretation. They show
that the Hamilton-Jacobi action $S(\mathbf{x},t)$ does not solve
only a given problem with a single initial condition $\left(
\mathbf{x}_{0}, \frac{\mathbf{\nabla }S_{0}\left(
\mathbf{x}_{0}\right) }{m}\right) $, but a set of problems with an
infinity of initial
conditions, all the pairs $\left( \mathbf{y},%
\frac{\mathbf{\nabla }S_{0}\left( \mathbf{y}\right) }{m}\right)$.
It answers the following question: "If we know the action (or the
velocity field) at the initial time, can we determine the action
(or the velocity field) at each later time?" This problem is
solved sequentially by the (local) evolution equation
(\ref{eq:HJ}). This is an \emph{a priori} point of view. It is the
problem solved by Nature with the principle of least action.

\begin{figure}[h]
\includegraphics[width=1\linewidth]{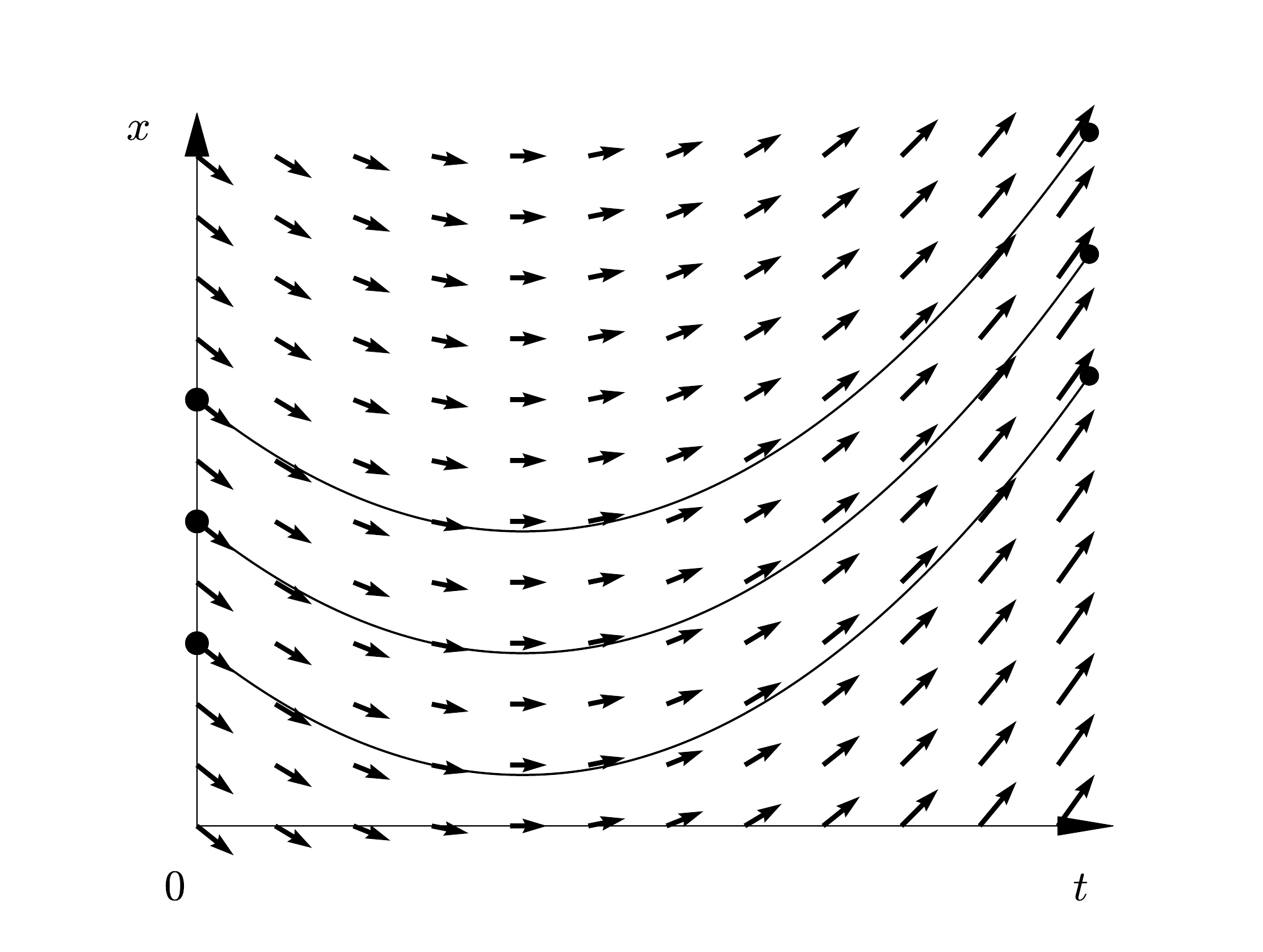}
\caption{\label{fig:fieldHJ} Velocity field that corresponds to the Hamilton-Jacobi action $ S\left( \mathbf{x},t\right)=m \textbf{v}_0 \cdot
\textbf{x} - \frac{1}{2} m \textbf{v}_0^2 t +\textbf{K}.\textbf{x}
t - \frac{1}{2} \textbf{K}.\textbf{v}_{0} t^{2} -
\frac{\textbf{K}^2 t^3}{6 m}$ ($\textbf{v}(\textbf{x},t)= \frac{\mathbf{\nabla }S\left( \mathbf{x},t\right)}{m}= \textbf{v}_0+ \frac{\textbf{K} t}{m}$) and three trajectories of particles piloted by this field.}
\end{figure}

For a particle in a linear potential $V(\textbf{x})= - \textbf{K}
.\textbf{x}$ with the initial action $S_0(\textbf{x})= m
\textbf{v}_0 \cdot \textbf{x}$, the initial velocity field is constant, 
$\textbf{v}(\textbf{x},0)= \frac{\mathbf{\nabla }S_{0}\left( \mathbf{x}\right)}{m}= \textbf{v}_0$ 
and the velocity field at time $t$ is also constant, 
$\textbf{v}(\textbf{x},t)= \frac{\mathbf{\nabla }S\left( \mathbf{x},t\right)}{m}= \textbf{v}_0+ \frac{\textbf{K} t}{m}$. 
Figure \ref{fig:fieldHJ} shows these velocity fields.

\section{Discerned and indiscerned particles in classical mechanics}
\label{sect:Discerned}

We show that the difficulties interpreting the action and the wave 
function result from the ambiguity in the definition of the conditions 
for the preparation of particles, which entails an ambiguity concerning 
the initial conditions. This ambiguity is related to the notion of 
indiscernibility which has never been well defined in the literature. 
It is responsible in particular for the Gibbs paradox: when calculating 
the entropy of a mixture of two identical gases in equilibrium, calculation 
by classical mechanics with distinguishable particles leads to an entropy twice as big as expected. If we replace these particles with indistinguishable 
particles, then the factor related to the indiscernibility yields the correct result.

In almost all textbooks on statistical mechanics, it is considered that this 
paradox stated by Willard Gibbs in 1889, was "solved" by quantum mechanics 
over thirty-five years later, thanks to the introduction of the indistinguishability 
postulate for identical quantum particles. Indeed, it was Einstein who, in 1924, 
introduced the indistinguishability of molecules of an ideal gas at the same time as the 
Bose-Einstein statistics. Nonetheless, as pointed out by Henri Bacry, "\emph{history might 
have followed a different path. Indeed, quite logically, we could have applied the principle 
of indiscernibility to save the Gibbs paradox. [...] This principle can be added to the 
postulates of quantum mechanics as well as to those of classical mechanics}". \cite{Bacry}

This same observation has been made by a large number
of other authors. In 1965 Land\'e~\cite{Lande_1965} demonstrated
that this indiscernability postulate of classical particles is
sufficient and necessary in order to explain why entropy vanished.
In 1977, Leinaas and Myrheim~\cite{Leinaas_1976} used it for the
foundation of their identical classical and quantum particles
theory. Moreover, as noted by Greiner et al.~\cite{Greiner_1999}, in addition to the Gibbs
paradox, several cases where it is needed to consider
indistinguishable particles in classical mechanics and
distinguishable particles in quantum mechanics can be found: "\textit{Hence, the Gibbs factor $
\frac{1}{N!}$ is indeed the correct recipe  for avoiding the Gibbs
paradox. From now on we will therefore always take into account
the Gibbs correction factor for indistinguishable states when we
count the microstates. However, we want to emphasize that this
factor is no more than a recipe to avoid the contradictions of
classical statistical mechanics. In the case of distinguishable
objects (e.g., atoms which are localized at certain grid points),
the Gibbs factor must not be added. In classical theory
the particles remain distinguishable. We will meet this
inconsistency more frequently in classical statistical
mechanics.}"~\cite{Greiner_1999} p.134  

We propose an accurate definition of both discernability
and indiscernability in classical mechanics and a way to avoid
ambiguities and paradoxes. Here, we only
consider the case of a single particle or a system of identical
particles without interactions and prepared in the same way. 

In classical mechanics, a particle is usually considered as a point and is described by its mass $m$, 
its charge if it has one, as well as its position $x_0$ and velocity $v_0$ at the initial instant. 
If the particle is subject to a potential field $V (x)$, we can deduce its path because its future 
evolution is given by Newton's or Lorentz's equations. This is why classical particles  are considered 
distinguishable. We will show, however, that a classical particle can be either non-discerned or 
discerned depending on how it is prepared.

We now consider a particle within a stationary beam of classical
identical particles such as electronic, atomic or molecular beams
($CO_2$ or $C_{60}$). At a very macroscopic level, one can consider a tennis ball canon.
Let us note that there is an abuse of language when one talks
about a classical particle. One should instead speak of a particle
that is studied in the framework of classical mechanics.

For a particle of this beam, we do not know at the initial instant the exact position 
or the exact velocity, only the characteristics describing the beam, that is to say, 
an initial probability density  $\rho _{0}\left(\mathbf{x}\right)$ and an initial velocity 
field $\mathbf{v}_0(\textbf{x})$ known through the initial action $S_0(\textbf{x})$ by the equation
$\mathbf{v}_0(\textbf{x})= \frac{\nabla S_0(\textbf{x})}{m}$ where
$m$ is the particle mass. This yields the following definition:

\begin{Definition}[Indiscerned prepared Particle]\label{defparticulesindiscernedenmc}- 
A classical particle is said to be indiscerned prepared 
when only the characteristics of the beam from which it comes (initial probability density 
$\rho _{0}\left( \textbf{x}\right)$ and initial action $S_0(\textbf{x}) $) are defined at the initial time.
\end{Definition}

In contrast, we have:

\begin{Definition}[Discerned prepared Particle]\label{defparticulesdiscernedenmc}- 
A classical particle is said to be discerned prepared, if one knows, 
at the initial time, its position $\textbf{x}_0 $ and velocity $\textbf{v}_0 $.
\end{Definition}

The notion of indiscernibility that we introduced does not depend on the observer's knowledge, 
but is related to the mode of preparation of the particle. 
 
Let us consider $N$ indiscerned particles, that is to say $N$ identical particles prepared 
in the same way, each with the same initial density $ \rho_{0}\left(\textbf{x}\right)$ and 
the same initial action $S_0(\textbf{x})$, subject to the same potential field $V(\textbf{x})$ 
and which will have independent behaviors. This is particularly the case of identical classical 
particles without mutual interaction and prepared the same way. It is also the case of identical 
classical particles such as electrons, prepared in the same way and which, although they may interact, 
will have independent behaviors if they are generated one by one in the system.

We called these particles indiscerned, and not indistinguishable, because if we knew their initial 
positions, their trajectories would be known. 

The difference between discerned and indiscerned particles depends on the preparation style.
A device prepares either discerned or indiscerned particles. 
By way of example a tennis ball machine randomly launches balls in different directions.
Therefore it prepares some indiscerned particles; only the characteristics of the balls' beams are known: probability of presence and velocity (action).
A tennis player plunged into complete darkness that uses this machine knows only the presence probability of balls. 
However %an indiscerned particule can became distinguishable if 
it is possible to discern indiscerned particles, if we knew their initial positions.
This is what happens during the day:
%This is what does a sighted tennis player; 
the tennis player is able to make successive measurements of the ball position by watching it.
In this case, the player is able to plan the trajectory.
It is important to note that without measurements, the balls remain indiscerned.
In this specific case, the position measurement changes neither the state of the particle nor its trajectories.
This is not always the case in quantum mechanics.
It is therefore easy for the observer to identify the indistinguishability of indiscerned particles.
However the tennis ball machine still produces indiscerned particles.
A shotgun that fires a number of small spherical pellets also produces a beam of indiscerned particules. 
The positions of the pellets are unknown, only their probability densities are known as well as their velocities.
If the precision of the shotgun is very high and if one uses a bullet (instead of pellets), 
the initial position $\textbf{x}_0$ of the bullet and its velocity $\textbf{v}_0$ are known with exactitude.
Therefore the bullet is a discerned prepared particle. 
The trajectories of the bullets will be always the same.
%The way whose are prepared the particlues is fundamental.
How the particles are prepared is fundamental.

Based on the previous definitions, we may state the following:
%1) A indiscerned particule whose initial position $x_0$ is also known is a discerned particule.
%2) A indiscerned particule whose initial probability density $\rho_0(x)$ is equal to a Dirac distribution $\rho_0(x)=\delta(x-x_0)$ is a discerned particule.
%Notice that following the pervious definitions:
\begin{enumerate}
 \item An indiscerned prepared particule whose initial position $x_0$ is also known is a discerned prepared particule.
 \item An indiscerned prepared particule whose initial probability density $\rho_0(x)$ is equal to a Dirac distribution $\rho_0(x)=\delta(x-x_0)$ is a discerned prepared particule.
\end{enumerate}
This means that the indiscerned particules can be distinguishable.  
Furthermore, in their enumerations indiscerned particules have the same 
properties that are usually granted to indistinguishable particles. Thus, if we select $N$ identical 
particles at random from the initial density $\rho_{0}\left( \textbf{x}\right)$, 
the various permutations of the $N$ particles are strictly equivalent 
and correspond, as for indistinguishable particles, to only one configuration. In this framework, 
the Gibbs paradox is no longer paradoxical as it applies to $N$ indiscerned particles whose different 
permutations correspond to the same configuration as for indistinguishable
particles. This means that if $X$ is the coordinate space of an
indiscerned particle, the true configuration space of $N$
indiscerned particles is not $X^N$ but rather $X^N/S_N $
where $S_N$ is the permutation group.

For indiscerned particles, we have the following theorem:

\begin{Theoreme}\label{th:eqstatHJ}- The probability density
$\rho \left( \mathbf{x},t\right)$ and the action $S\left(
\mathbf{x,}t\right)$ of classical particles prepared in the same
way, with initial density $\rho_0( \textbf{x})$, with the same
initial action $S_0(\textbf{x})$, and evolving in the same
potential $V(\textbf{x})$, are solutions to the
\textbf{statistical Hamilton-Jacobi equations}:
\begin{eqnarray}\label{eq:statHJ1b}
\frac{\partial S\left(\textbf{x},t\right) }{\partial
t}+\frac{1}{2m}(\nabla S(\textbf{x},t) )^{2}+V(\textbf{x})=0\\
\label{eq:statHJ2b}
S(\textbf{x},0)=S_{0}(\textbf{x})\\
\label{eq:statHJ3b}
\frac{\partial \mathcal{\rho }\left(\textbf{x},t\right) }{\partial
t}+ div \left( \rho \left( \textbf{x},t\right) \frac{\nabla
S\left( \textbf{x},t\right) }{m}\right) =0\\
\label{eq:statHJ4b}
\rho(\mathbf{x},0)=\rho_{0}(\mathbf{x}).
\end{eqnarray}
\end{Theoreme}

Let us recall that the velocity field is
$\textbf{v}(\textbf{x},t)= \frac{\nabla S\left(
\textbf{x},t\right) }{m}$ and that the Hamilton-Jacobi equation
(\ref{eq:statHJ1b}) is not coupled to the continuity equation
(\ref{eq:statHJ3b}).

The difference between discerned and indiscerned particles will 
provide a simple explanation to the "recipes" denounced by Greiner et al.~\cite{Greiner_1999} 
that are commonly presented in manuals on classical statistical mechanics. 
However, as we have seen, this is not a principle that must be added. 
The nature of this discernability of the particle depends on the preparation 
conditions of the particles, whether discerned or indiscerned.

Can we define an action for a discerned particle in a potential field $V (\mathbf{x})$? 
Such an action should depend only on the starting point $\mathbf{x}_0 $, 
the initial velocity $\mathbf{v}_0$ and the potential field $V (\mathbf{x})$.

\begin{Theoreme}\label{th:actionponctuelle}- If $\xi(t)$
is the classical trajectory in the field $V(\textbf{x})$ of a
particle with the initial position $\textbf{x}_0$ and with initial
velocity $\textbf{v}_0$, then the function
\begin{equation}\label{eq:soleqHJponctuelle}
S\left( \mathbf{x},t;\textbf{x}_0,\textbf{v}_0 \right)= m \frac{d \xi(t)}{dt} \cdot
\textbf{x} + g(t)
\end{equation}
where $\frac{d g(t)}{dt}= -\frac{1}{2}m (\frac{d \xi(t)}{dt})^2-
V(\xi(t)) - m \frac{d^2 \xi(t)}{dt^2} \cdot \xi(t)$, is called
\textbf{the deterministic action}, and is a solution to deterministic Hamilton-Jacobi
equations:
\begin{eqnarray}
0=\frac{\partial S\left(\textbf{x},t;\textbf{x}_0,\textbf{v}_0\right) }{\partial
t}|_{\textbf{x}=\xi(t)}
&+&\frac{1}{2m}(\nabla S(\textbf{x},t;\textbf{x}_0,\textbf{v}_0))^{2}|_{\textbf{x}=\xi(t)}\nonumber\\
&+&V(\textbf{x})|_{\textbf{x}=\xi(t)}\label{eq:statHJponctuelle1b}
\end{eqnarray}
\begin{eqnarray}\label{eq:statHJponctuelle1c}
\frac{d\xi(t)}{dt}=\frac{\nabla S(\xi(t),t;\textbf{x}_0,\textbf{v}_0)}{m}\\
\label{eq:statHJponctuelle1d}
S(\textbf{x},0;\textbf{x}_0,\textbf{v}_0)= m \textbf{v}_0 \textbf{x}
~\text{ and }~\xi(0)=\textbf{x}_0.
\end{eqnarray}
\end{Theoreme}

The deterministic action $ S(\textbf{x},t;\textbf{x}_0,\textbf{v}_0)$ satisfies the 
Hamilton-Jacobi equations only along the trajectory $\xi(t)$. The interest of such 
an action related to a single localized trajectory is above all theoretical by 
proposing a mathematical framework for the discerned particle. This action will take
on a meaning in the following section where we show that it corresponds to the limit 
of the wave function of a quantum particle in a coherent state when one makes the 
Planck constant $h$ tend to $0$.

As for the Hamilton-Jacobi action, the deterministic action only depends on 
the initial conditions ($\textbf{x}_0,\textbf{v}_0 $), the "efficient causes". 
In the end, we have three actions in classical mechanics, an epistemological action (the Euler-Lagrange action $ S(\textbf{x},t;\textbf{x}_0)$) and two ontological actions, 
the Hamilton-Jacobi action $ S(\textbf{x},t)$ for the indiscerned particles and 
the deterministic action $ S(\textbf{x},t;\textbf{x}_0,\textbf{v}_0)$ for the discerned particles.

\section{The two limits of the Schr\"odinger equation}
\label{sect:2limits}

Let us consider the case semi-classical where the wave function $\Psi(\textbf{x},t)$ is a solution to the Schr\"odinger
equation :
\begin{eqnarray}\label{eq:schrodinger1}
i\hbar \frac{\partial \Psi }{\partial t}=\mathcal{-}\frac{\hbar ^{2}}{2m}%
\triangle \Psi +V(\mathbf{x},t)\Psi\\
\label{eq:schrodinger2}
\Psi (\mathbf{x},0)=\Psi_{0}(\mathbf{x}).
\end{eqnarray}
With the variable change $\Psi(\mathbf{x},t)=\sqrt{\rho^{\hbar}(\mathbf{x},t)} \exp(i\frac{S^{\hbar}(\textbf{x},t)}{\hbar})$, the Schr\"odinger
equation can be decomposed into Madelung
equations~\cite{Madelung_1926} (1926):
\begin{equation}\label{eq:Madelung1}
\frac{\partial S^{\hbar}(\mathbf{x},t)}{\partial t}+\frac{1}{2m}
(\nabla S^{\hbar}(\mathbf{x},t))^2 +
V(\mathbf{x},t)-\frac{\hbar^2}{2m}\frac{\triangle
\sqrt{\rho^{\hbar}(\mathbf{x},t)}}{\sqrt{\rho^{\hbar}(\mathbf{x},t)}}=0
\end{equation}
\begin{equation}\label{eq:Madelung2}
\frac{\partial \rho^{\hbar}(\mathbf{x},t)}{\partial t}+ div
\left(\rho^{\hbar}(\mathbf{x},t) \frac{\nabla
S^{\hbar}(\mathbf{x},t)}{m}\right)=0
\end{equation}
with initial conditions:
\begin{equation}\label{eq:Madelung3}
\rho^{\hbar}(\mathbf{x},0)=\rho^{\hbar}_{0}(\mathbf{x}) \qquad \text{and}
\qquad S^{\hbar}(\mathbf{x},0)=S^{\hbar}_{0}(\mathbf{x}) .
\end{equation}
We consider two cases \textit{depending on the preparation of the
particles}~\cite{Gondran2011,Gondran2012a}.

\begin{Definition}[Semi-Classical indiscerned particle]\label{defdensiteinitstat}
- A quantum particle is said to be semi-classical indiscerned
prepared if its initial probability density $\rho^{\hbar}_{0}(\mathbf{x})$
and its initial action $S^{\hbar}_{0}(\mathbf{x})$ are regular
functions $\rho_{0}(\mathbf{x})$ and $S_{0}(\mathbf{x})$ not
depending on $\hbar$.
\end{Definition}

It is the case of a set of non-interacting particles all prepared
in the same way: a free particle beam in a linear potential, an
electronic or $C_{60}$ beam in the Young's slits diffraction, or
an atomic beam in the Stern and Gerlach experiment.
\begin{Definition}[Semi-Classical discerned particle]\label{defdensiteinitponct}- 
A quantum particle is said to be semi-classical discerned prepared
if its initial probability density $\rho^{\hbar}_{0}(\mathbf{x})$
converges, when $\hbar\to 0$, to a Dirac distribution and if its
initial action $S^{\hbar}_{0}(\mathbf{x})$ is a regular function
$S_{0}(\mathbf{x})$ not depending on $\hbar$.
\end{Definition}
This situation occurs when the wave packet corresponds to a
quasi-classical coherent state, introduced in 1926 by
Schr\"odinger~\cite{Schrodinger_26}. The field quantum theory and
the second quantification are built on these coherent
states~\cite{Glauber_65}. The existence for the hydrogen atom of a
localized wave packet whose motion is on the classical trajectory
(an old dream of Schr\"odinger's) was predicted in 1994 by
Bialynicki-Birula, Kalinski, Eberly, Buchleitner and
Delande~\cite{Bialynicki_1994, Delande_1995, Delande_2002}, and
discovered recently by Maeda and Gallagher~\cite{Gallagher} on
Rydberg atoms.

\subsection{Semi-Classical indiscerned quantum particles}

\begin{Theoreme}~\cite{Gondran2011,Gondran2012a} For semi-classical indiscerned quantum particles,
the probability density $\rho^{\hbar}(\textbf{x},t)$ and the
action $S^{\hbar}(\textbf{x},t)$, solutions to the Madelung
equations
(\ref{eq:Madelung1})(\ref{eq:Madelung2})(\ref{eq:Madelung3}),
converge, when $\hbar\to 0$, to the classical density
$\rho(\textbf{x},t)$ and the classical action $S(\textbf{x},t)$,
solutions to the statistical Hamilton-Jacobi equations (\ref{eq:statHJ1b})(\ref{eq:statHJ2b})(\ref{eq:statHJ3b})(\ref{eq:statHJ4b}).
\end{Theoreme}

We give some indications on the demonstration of this theorem and
we propose its interpretation. Let us consider the case where the
wave function $\Psi(\textbf{x},t)$ at time $t$ is written as a
function of the initial wave function $\Psi_{0}(\textbf{x})$ by
the Feynman paths integral \cite{Feynman_1965} (\ref{eq:interFeynman}). For a semi-classical indiscerned quantum particle, the wave function is written
$ \Psi(\textbf{x},t)= F(t,\hbar)\int\sqrt{\rho_0(\mathbf{x}_0)}
\exp(\frac{i}{\hbar}( S_0(\textbf{x}_0)+
S_{cl}(\textbf{x},t;\textbf{x}_{0})) d\textbf{x}_0$. The theorem
of the stationary phase shows that, if $\hbar$ tends towards 0, we
have $ \Psi(\textbf{x},t)\sim
\exp(\frac{i}{\hbar}min_{\textbf{x}_0}( S_0(\textbf{x}_0)+
S_{cl}(\textbf{x},t;\textbf{x}_{0}))$, that is to say that the
quantum action $S^{h}(\textbf{x},t)$ converges to the function
\begin{equation}\label{eq:solHJminplus}
S(\textbf{x},t)=min_{\textbf{x}_0}( S_0(\textbf{x}_0)+
S_{cl}(\textbf{x},t;\textbf{x}_{0}))
\end{equation}
which is the solution to the Hamilton-Jacobi equation
(\ref{eq:statHJ1b}) with the initial condition (\ref{eq:statHJ2b}).
Moreover, as the quantum density $\rho^{h}(\textbf{x},t)$
satisfies the continuity equation (\ref{eq:Madelung2}), we deduce,
since $S^{h}(\textbf{x},t)$ tends towards $S(\textbf{x},t)$, that
$\rho^{h}(x,t)$ converges to the classical density
$\rho(\textbf{x},t)$, which satisfies the continuity equation
(\ref{eq:statHJ3b}). We obtain both announced convergences.

\begin{figure}[h]
\includegraphics[width=0.9\linewidth]{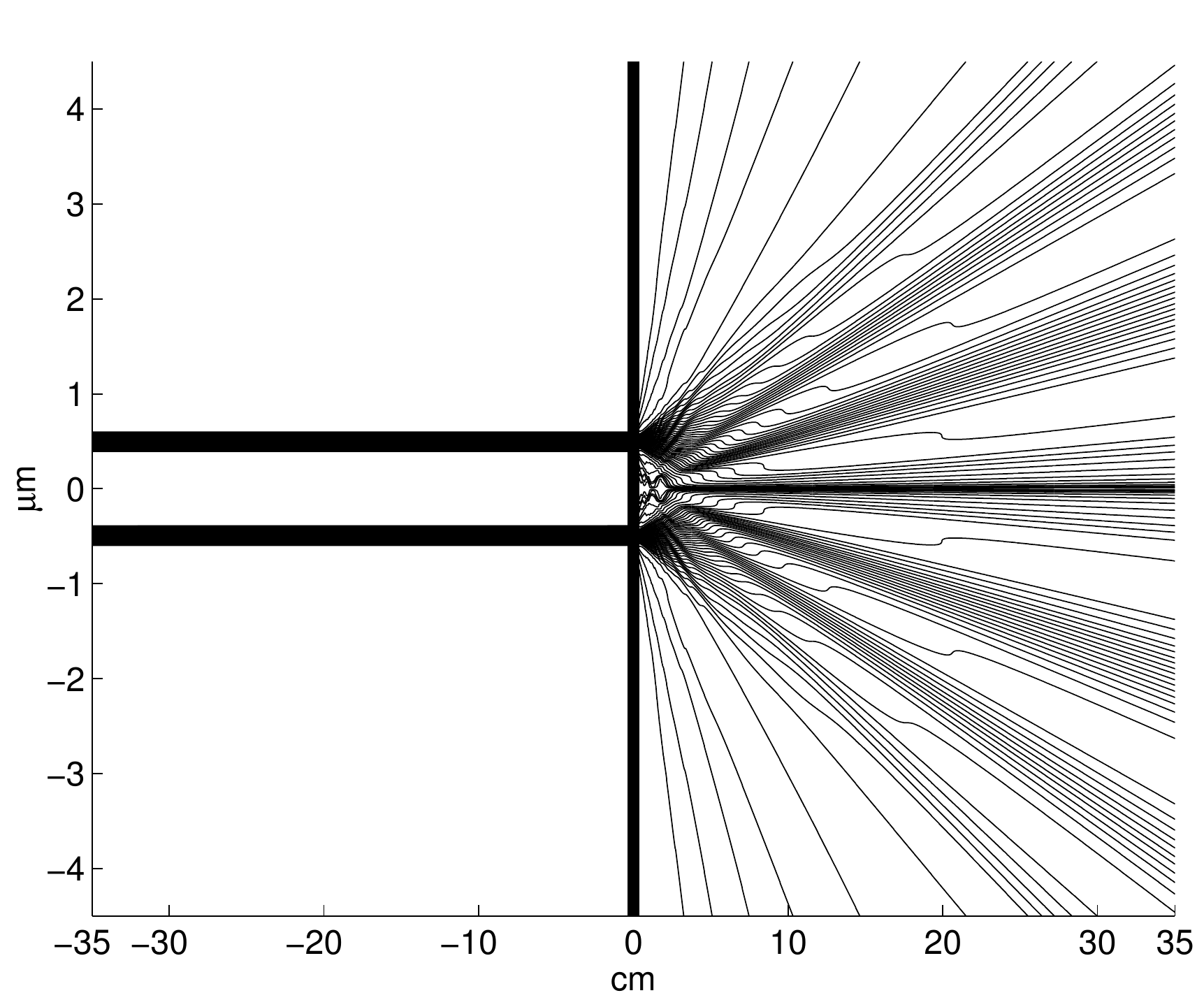}
\caption{\label{fig:traj-Young} 100 electron trajectories for the J\"onsson experiment.}
\end{figure}

For a semi-classical indiscerned quantum particle, the Madelung equations converge to 
the statistical Hamilton-Jacobi equations, which correspond to indiscerned 
classical particles. We use now the interpretation of the statistical Hamilton-Jacobi 
equations to deduce the interpretation of the Madelung equations. For these indiscerned 
classical particles, the density and the action are not sufficient to describe a classical particle. 
To know its position at time $t$, it is necessary to know its initial position. 
It is logical to do the same in quantum mechanics. We consider this indiscerned quantum particle as the classical particle.

We conclude that a \textit{semi-classical indiscerned quantum particle} is not completely described by
its wave function. It is necessary  to add its initial
position and it becomes natural to introduce the de Broglie-Bohm interpretation \cite{Broglie1927,Bohm_52}. 
In this interpretation, the two first postulates of quantum mechanics, describing the
quantum state and its evolution, must be completed. 
At initial time $t=0$, the state of the particle is
given by the initial wave function $ \Psi_{0}(\textbf{x})$ (a wave
packet) and its initial position $\textbf{X}(0)$; it is the new first postulate.
The second new postulate gives the evolution on the wave function and on the position. 
For a single, spin-less particle in a potential
$V(\textbf{x})$, the evolution of the wave function is given by the usual
Schr\"odinger equation (\ref{eq:schrodinger1})(\ref{eq:schrodinger2}) and the evolution of the particle position is given by
\begin{equation}\label{eq:champvitesse}
\frac{d
\textbf{X}(t)}{dt}=\frac{1}{m}\nabla
S^{\hbar}(\textbf{x},t)|_{\textbf{x}=\textbf{X}(t)}.
\end{equation}

In the case of a particle with spin, as in the Stern and Gerlach
experiment, the Schr\"odinger equation must be replaced by the
Pauli or Dirac equations.

The other quantum mechanics postulates which describe the measurement 
are not necessary. One can demonstrate that
the three postulates of measurement can be explained on each
example; see the double-slit, Stern-Gerlach and EPR-B  experiments\cite{Gondran_2013}. These postulates are remplaced by a single one, the "quantum equilibrium 
hypothesis", that describes the interaction 
between the initial wave function $\Psi_0(\textbf{x})$ and the probabilty distribution of the initial particle position
$\textbf{X}(0)$:
\begin{equation}\label{eq:quantumequi}
P[\textbf{X}(0)=\textbf{x}]=|\Psi_0(\textbf{x})|^2 .
\end{equation}
One deduces that for all times
\begin{equation}\label{eq:quantumequit}
P[\textbf{X}(t)=\textbf{x}]=|\Psi(\textbf{x},t)|^2.
\end{equation}
This is the "equivariance" property of the
$|\Psi(\textbf{x},t)|^2$ probability distribution~\cite{Durr_1992}
which yields the Born probabilistic interpretation.
Let us note the minimality of the de Broglie-Bohm interpretation.

Figure~\ref{fig:traj-Young} shows a simulation of the de Broglie-Bohm trajectories 
in the double slit experiment of J\"onsson \cite{Jonsson_1961} where an electron 
gun emits electrons one by one through a hole with a radius of a few micrometers. 
The electrons, prepared similarly, are represented by the same initial wave function, 
but not by the same initial position. In the simulation, these initial positions are randomly selected in the
initial wave packet. We have represented only the quantum trajectories through one of two slits.

\begin{figure}[h]
\includegraphics[width=0.9\linewidth]{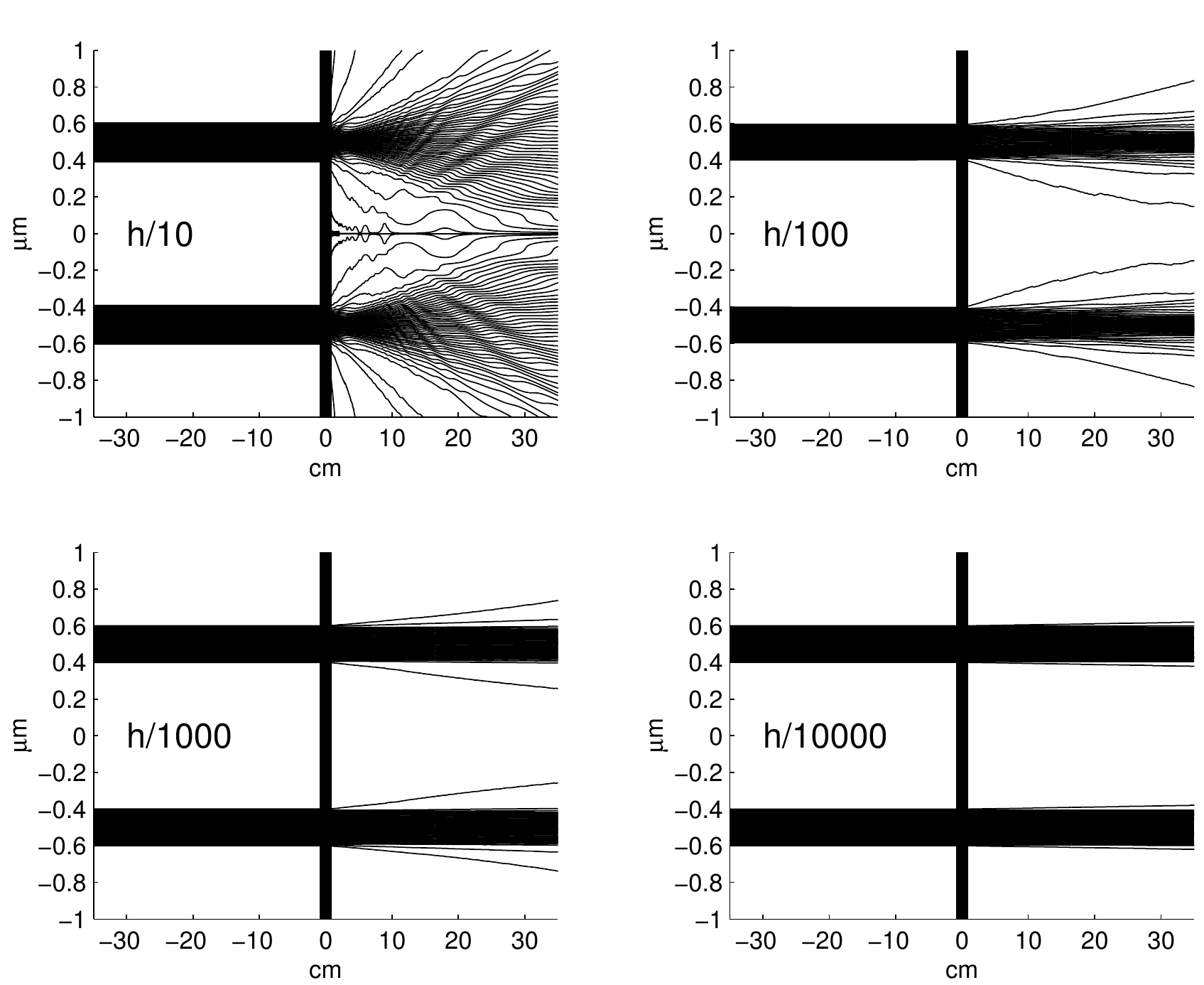}
\caption{\label{fig:traj-Young-converg}.Convergence of 100 electron trajectories when h is divided by 10, 100, 1 000 and 10 000.}
\end{figure}

Figure~\ref{fig:traj-Young-converg} shows the 100 previous trajectories when 
the Planck constant is divided by 10, 100, 1000 and 10000 respectively. 
We obtain, when h tends to 0, the convergence of quantum trajectories to classical trajectories.

\subsection{Semi-Classical discerned quantum particles}

The convergence study of the semi-classical discerned quantum particle is
mathematically very difficult. We only study the example of a
coherent state where an explicit calculation is possible.

For the two dimensional harmonic oscillator,
$V(\textbf{x})=\frac{1}{2}m \omega^{2}\textbf{x}^{2}$, coherent
states are built~\cite{CT_1977} from the initial wave
function $\Psi_{0}(\textbf{x})$ which corresponds to the density
and initial action $ \rho^{\hbar}_{0}(\mathbf{x})= ( 2\pi \sigma
_{\hbar}^{2}) ^{-1} e^{-\frac{( \textbf{x}-\textbf{x}_{0})
^{2}}{2\sigma _{\hbar}^{2}}}$ and $
S_{0}(\mathbf{x})=S^{\hbar}_{0}(\mathbf{x})= m \textbf{v}_{0}\cdot
\textbf{x}$ with $ \sigma_\hbar=\sqrt{\frac{\hbar}{2 m \omega}}$.
Here, $\textbf{v}_0$ and $\textbf{x}_0$ are constant vectors
and independent from $\hbar$, but $\sigma_\hbar$ will tend to $0$
as $\hbar$. With initial conditions, the density
$\rho^{\hbar}(\textbf{x},t)$ and the action
$S^{\hbar}(\textbf{x},t)$, solutions to the Madelung equations
(\ref{eq:Madelung1})(\ref{eq:Madelung2})(\ref{eq:Madelung3}), are
equal to ~\cite{CT_1977}:
$\rho^{\hbar}(\textbf{x},t)=\left( 2\pi \sigma_{\hbar} ^{2}
\right) ^{-1}e^{- \frac{( \textbf{x}-\xi(t)) ^{2}}{2\sigma_{\hbar}
^{2} }}$ and $S^{\hbar}(\textbf{x},t)= + m \frac{d\xi
(t)}{dt}\cdot \textbf{x} + g(t) - \hbar\omega t$, where $\xi(t)$
is the trajectory of a classical particle evolving in the
potential $V(\textbf{x})=\frac{1}{2} m \omega^{2} \textbf{x}^2 $,
with $\textbf{x}_0$ and $\textbf{v}_0$ as initial position and
velocity and $g(t)=\int _0 ^t ( -\frac{1}{2} m (\frac{d\xi
(s)}{ds})^{2} + \frac{1}{2} m \omega^{2} \xi(s)^2) ds$.

\begin{Theoreme}\label{t-convergenceparticulediscerne}~\cite{Gondran2011,Gondran2012a}- 
For the harmonic oscillator, when $\hbar\to 0$,
the density $\rho^{\hbar}(\textbf{x},t)$ and the action
$S^{\hbar}(\textbf{x},t)$ converge to
\begin{equation}
\rho(\textbf{x},t)=\delta( \textbf{x}- \xi(t)) ~\text{ and }~
S(\textbf{x},t)= m \frac{d\xi (t)}{dt}\cdot\textbf{x} + g(t)
\end{equation}
where $S(\textbf{x},t)$ and the trajectory $\xi(t)$ are solutions
to the deterministic Hamilton-Jacobi equations (\ref{eq:statHJponctuelle1b})(\ref{eq:statHJponctuelle1c})(\ref{eq:statHJponctuelle1d}).
\end{Theoreme}

Therefore, the kinematic of the wave packet converges to the
single harmonic oscillator described by $\xi(t)$, which corresponds to \emph{a
discerned classical particle}. It is then possible to consider, unlike for the semi-classical indiscerned particles, that the wave function can be
viewed as a single quantum particle. Then, we consider this discerned quantum particle as the classical particle. 
The \textit{semi-classical discerned quantum particle} is in line with the Copenhagen interpretation
of the wave function, which contains all the information on the
particle. A natural interpretation was proposed by
Schr\"odinger~\cite{Schrodinger_26} in 1926 for the coherent
states of the harmonic oscillator: the quantum particle is a
spatially extended particle, represented by a wave packet whose
center follows a classical trajectory. In this interpretation, 
the first two usual postulates of quantum mechanics are maintened. 
The others are not necessary. 
Then, the particle center is the mean value of the position 
($X(t)=\int x\vert\Psi(\textbf{x},t) \vert^2 dx$)  and satisfies the Ehrenfest theorem~\cite{Ehrenfest}. 
Let us note the minimality of the Schr\"odinger interpretation.

\section{The non semi-classical case}
\label{sect:NonSemiClassical}

The de Broglie-Bohm and Schr\"odinger interpretations correspond to the
semi-classical approximation.  They correspond to the two
interpretations proposed in 1927 at the Solvay congress~\cite{Solvay} by de
Broglie and Schr\"odinger. But there exist situations in which the
semi-classical approximation is not valid. It is in particular the
case of state transitions in a hydrogen atom. Indeed, since
Delmelt'experiment~\cite{Delmelt_1986} in 1986, the physical
reality of individual quantum jumps has been fully validated. The
semi-classical approximation, where the interaction with the
potential field can be described classically, is no longer
possible and it is necessary to use electromagnetic field 
quantization since the exchanges occur photon by photon. In this situation, the
Schr\"odinger equation cannot give a deterministic interpretation
and the statistical Born interpretation is the only valid one. It
was the third interpretation proposed in 1927 at the Solvay
congress. These three interpretations are considered by Einstein in one of
his last articles (1953), "\textit{Elementary Reflexion on
Interpreting the Foundations of Quantum Mechanics}" in a homage to
Max Born~\cite{Einstein}:

"\textit{The fact that the Schr\"odinger equation associated with
the Born interpretation does not lead to a description of the
"real states" of an individual system, naturally incites one to
find a theory that is not subject to this limitation.} \textit{Up
to now, the two attempts have in common that they conserve the
Schr\"odinger equation and abandon the Born interpretation.}
\emph{The first one, which marks de Broglie's comeback, was
continued by Bohm.... The second one, which aimed to get a "real
description" of an individual system and which might be based on
the Schr\"odinger equation is very late and is from Schr\"odinger
himself. The general idea is briefly the following : the function
$\psi$ represents in itself the reality and it is not necessary to
add it to Born's statistical interpretation.}[...] \textit{From
previous considerations, it results that the only acceptable
interpretation of the Schr\"odinger equation is the statistical
interpretation given by Born. Nevertheless, this interpretation
doesn't give the "real description" of an individual system, it
just gives statistical statements of entire systems.}"

Thus, Einstein retained de Broglie's and Schr\"odinger's attempts to 
interpret the "real states" of a single system: these are our 
two semi-classical quantum particles (indiscerned and discerned). 
But as de Broglie and Schr\"odinger retained the Schr\"odinger equation, 
Einstein, who considered this equation as fundamentally statistical, 
refuted each of their interpretations. We will see that he went too far in his rebuttal.

The novelty of our approach is to consider that each of these three 
interpretations depends on the preparation of the particles. 
The de Broglie-Bohm interpretation concerns the semi-classical indiscerned quantum 
particles, the Schr\"odinger interpretation concerns 
the semi-classical discerned quantum particles, and the Born interpretation 
concerns the statistic of the semi-classical indiscerned quantum particles, 
but also the statistic of the transitions in a hydrogen atom.

This does not mean that we should abandon determinism and realism, but that at this scale, 
Schr\"odinger's statistical wave function is not the effective equation  to obtain an individual
behavior of a particle, in particular to investigate the instant of 
transition in a deterministic manner. An individual interpretation needs to use
the creation and annihilation operators of quantum field
theory, but this interpretation still remains statistical.

We assume that it is possible to construct a deterministic
quantum field theory that extends to the non semi-classical
interpretation of the double semi-classical preparation. First,
as shown by de Muynck~\cite{Muynck}, we can construct a theory
with discerned (labeled) creation and annihilation operators in
addition to the usual indiscerned creation and annihilation
operators. But, to satisfy the determinism, it is necessary to
search, at a lower scale, the mechanisms that allow the emergence of
the creation operator.

\section{Conclusion}
\label{sect:conclusion}

The introduction into classical mechanics of the concepts of indiscerned 
particles verifying the statistical Hamilton-Jacobi equations and of discerned 
particles verifying the deterministic Hamilton-Jacobi equations can provide a 
simple answer to the Gibbs paradox of classical statistical mechanics. 
Furthermore, the distinction between the Hamilton-Jacobi and Euler-Lagrange actions, 
based on the Minplus path integral makes it easier to understand the principle of least action.
The study of the convergence of the Madelung equations when $h$ tends to $0$ leads us to consider the following two cases:
\begin{enumerate}
 \item Semi-classical indiscerned quantum particles, which are prepared in the same way and without mutual interaction, for which the evolution equations converge to the statistical 
Hamilton-Jacobi equations of indiscerned classical particles. The wave function is therefore not 
sufficient to represent quantum particles and it is mandatory to add their initial positions, 
just as for indiscerned classical particles. Subsequently, the interpretation of the de Broglie-Bohm pilot wave is  necessary.
 \item The semi-classical discerned quantum particle for which the evolution equations 
converge to the deterministic Hamilton-Jacobi equations of a discerned classical particle. 
The interpretation of the Broglie-Bohm pilot wave is no longer necessary because the wave 
function is sufficient to represent the particles as in the Copenhagen interpretation. 
Subsequently, the Schr\"odinger 
interpretation where the wave function represents an extended particle, is the most natural. 
\end{enumerate}
We can consider the previous interpretation which depends on a double preparation of the quantum particle (discerned or non-discerned)  
as a response to the "theory of the double solution" that Louis de Broglie was seeking in 1927. We call it "the theory of the double preparation".

In the case where the semi-classical approximation is no longer
valid, the interpretation needs to use
the creation and annihilation operators of the quantum field.
W. M. de Muynck~\cite{Muynck} shows that is possible to construct a theory
with discerned (labeled) creation and annihilation operators in
addition to the usual non-discerned creation and annihilation
operators. But, to satisfy the determinism, it is necessary to
search, at a lower scale, the mechanisms that allow the emergence of
the creation operator.

This interpretation of quantum mechanics following the preparation of the system can explain the discussions 
between the founding fathers, in particular the discussion of the Solvay Congress of 1927. 
Indeed, one may consider that the misunderstanding between them may have come from the fact that they each had 
an element of truth: Louis de Broglie's pilot-wave interpretation for the semi-classical indiscerned  particle, 
Schr\"odinger's interpretation for the semi-classical discerned particle and Born's interpretation for 
the non-semi-classical case. But each applied his particular case to the general case and they consequently 
made mutually incompatible interpretations.

\end{document}